\documentclass{article}
\usepackage{booktabs,subcaption,amsfonts,dcolumn}

\usepackage[english]{babel}

\usepackage[letterpaper,top=2cm,bottom=2cm,left=3cm,right=3cm,marginparwidth=1.75cm]{geometry}

\usepackage{amsmath}
\usepackage{graphicx}
\usepackage{xcolor}
\usepackage{url}
\usepackage{bm}
\usepackage[colorlinks=true, allcolors=blue]{hyperref}

\newcommand{\Gate}[1]{\textsc{#1}}

\newcommand{\zgate}{\Gate{z}}

\newcommand{\xgate}{\Gate{x}}

\newcommand{\idgate}{\Gate{i}}

\newcommand{\cnotgate}{\Gate{cnot}}

\newcommand*{\Prob}{\mathsf{Pr}}

\title{QAOA with $N\cdot p\geq 200$}
\author{Ruslan Shaydulin and Marco Pistoia \\ {\small Global Technology Applied Research, JPMorgan Chase, New York}}
\date{}

\begin{document}
\maketitle

\begin{abstract}
One of the central goals of the DARPA Optimization with Noisy Intermediate-Scale Quantum (ONISQ) program is to implement a hybrid quantum/classical optimization algorithm with high $N\cdot p$, where $N$ is the number of qubits and $p$ is the number of alternating applications of parameterized quantum operators in the protocol. In this note, we demonstrate the execution of the Quantum Approximate Optimization Algorithm (QAOA) applied to the MaxCut problem on non-planar 3-regular graphs with $N\cdot p$ of up to $320$ on the Quantinuum H1-1 and H2 trapped-ion quantum processors. To the best of our knowledge, this is the highest $N\cdot p$ demonstrated on hardware to date. Our demonstration highlights the rapid progress of quantum hardware.
\end{abstract}

\section*{Introduction}

The Quantum Approximate Optimization Algorithm (QAOA)~\cite{Hogg2000,farhi2014quantum} is one of the leading candidate algorithms for demonstrating better-than-classical performance on near-term quantum computers. As a consequence, the task of implementing QAOA on hardware has attracted a lot of attention. For example, the central goal of Technical Area 1 of the DARPA ONISQ program is to implement a quantum optimizer with high $N\cdot p$~\cite{DARPA_ONISQ_BAA}, with $N\cdot p > 100$ as the target for Phase 1, and $N\cdot p > 10,000$ for Phase 2. Here, $N$ refers to the number of qubits used, while $p$ is the number of alternating applications of QAOA operators (commonly referred to as QAOA layers). QAOA has also been proposed as a scalable application-centric benchmark for quantum hardware~\cite{Martiel2021}.

In this note, we report two sets of experiments. First, we report a successful execution of QAOA with classically-optimized parameters applied to the MaxCut problem on 3-regular graphs with $N = 20$ and $p \geq 10$ on the Quantinuum H1-1 trapped-ion quantum processor. We consider an execution with $p$ layers successful if, for all $1\leq p' \leq p$, the solution quality at $p'$ is greater than at $p'-1$. For all $N=20$ problem instances considered, we observe that the solution quality obtained by the algorithm increases monotonically with $p$ up to $10$. For some instances we observe monotonic improvements for $p$ as large as $15$. Second, we extend the experiments to $N=32$ by using QAOA with fixed parameters~\cite{wurtz2021fixedangle} on the Quantinuum H-2 processor. Due to parameters not being optimized on per-instance basis, we observe less consistent QAOA performance. Still, at $N=32$ we observe monotonic improvement of QAOA performance with $p$ up to $10$, giving $N\cdot p = 320$. To the best of our knowledge, this is the largest (in terms of $N\cdot p$) QAOA demonstration on gate-model quantum computers to date. We make the executed circuits along with the raw data obtained from hardware publicly available at \url{https://doi.org/10.5281/zenodo.8338585}.

\section*{Results}

\paragraph{Problem Definition} We consider the MaxCut problem on 3-regular graphs. The goal of MaxCut is to partition the set of nodes $V$ of a graph $G=(V,E)$ into two disjoint subsets such that the number of edges in $E$ spanning both parts is maximized. For a sequence of spins $z\in \{-1,1\}^{|V|}$, the MaxCut objective is given by $\mathcal{C}(z) = \frac{1}{2}\sum_{( i,j )\in E} \left(1-z_i z_j\right)$. This objective is encoded on qubits by a diagonal Hamiltonian $C = \frac{1}{2}\sum_{( i,j )\in E} \left(\idgate-\zgate_i\zgate_j\right)$,
where $\zgate_k$ is a Pauli $\zgate$ acting on qubit $k$.

The quantum state prepared by QAOA circuit with $p$ layers is then given by 
\[
\vert \bm \beta, \bm \gamma\rangle = \prod_{l=1}^p e^{-i \beta_l \sum_{j=1}^{|V|}\xgate_j} e^{-i \gamma_l C} \vert + \rangle^{\otimes |V|}, 
\]
where $\vert + \rangle^{\otimes |V|}$ is a uniform superposition over computational basis states and $\xgate_j$ is a Pauli $\xgate$ acting on qubit $j$. The central figures of merit are the expected objective value (expected cut in the case of MaxCut), given by
\[
\langle C\rangle =  \langle \bm \beta, \bm \gamma \vert C\vert \bm \beta, \bm \gamma\rangle = \sum_{z\in \{0,1\}^{|V|}}\Prob(z)\mathcal{C}(z),
\]
and the probability of obtaining the optimal solution when measuring the QAOA state, denoted as
$p^{\text{opt}}$. For some instances, we additionally report approximation ratio, given by $\langle C\rangle / \mathcal{C}^*$, where $\mathcal{C}^* = \max_{z\in \{-1,1\}^{|V|}}\mathcal{C}(z)$ is the optimal value of the cut. As can be easily seen, if the parameters are optimized with respect to a given metric, the correspondingly defined solution quality of  QAOA can only increase with the number of layers $p$ in the absence of noise. A noisy device introduces a trade-off between the improvements in solution quality and the increased probability of error from adding more layers. This means that, in practice, there is a depth beyond which adding more layers is not beneficial. In general, this depth is higher if the error rates are lower, motivating the use of $N\cdot p$ as an application-centric measure of device performance.

\begin{figure*}
    \centering
    \includegraphics[width=\textwidth]{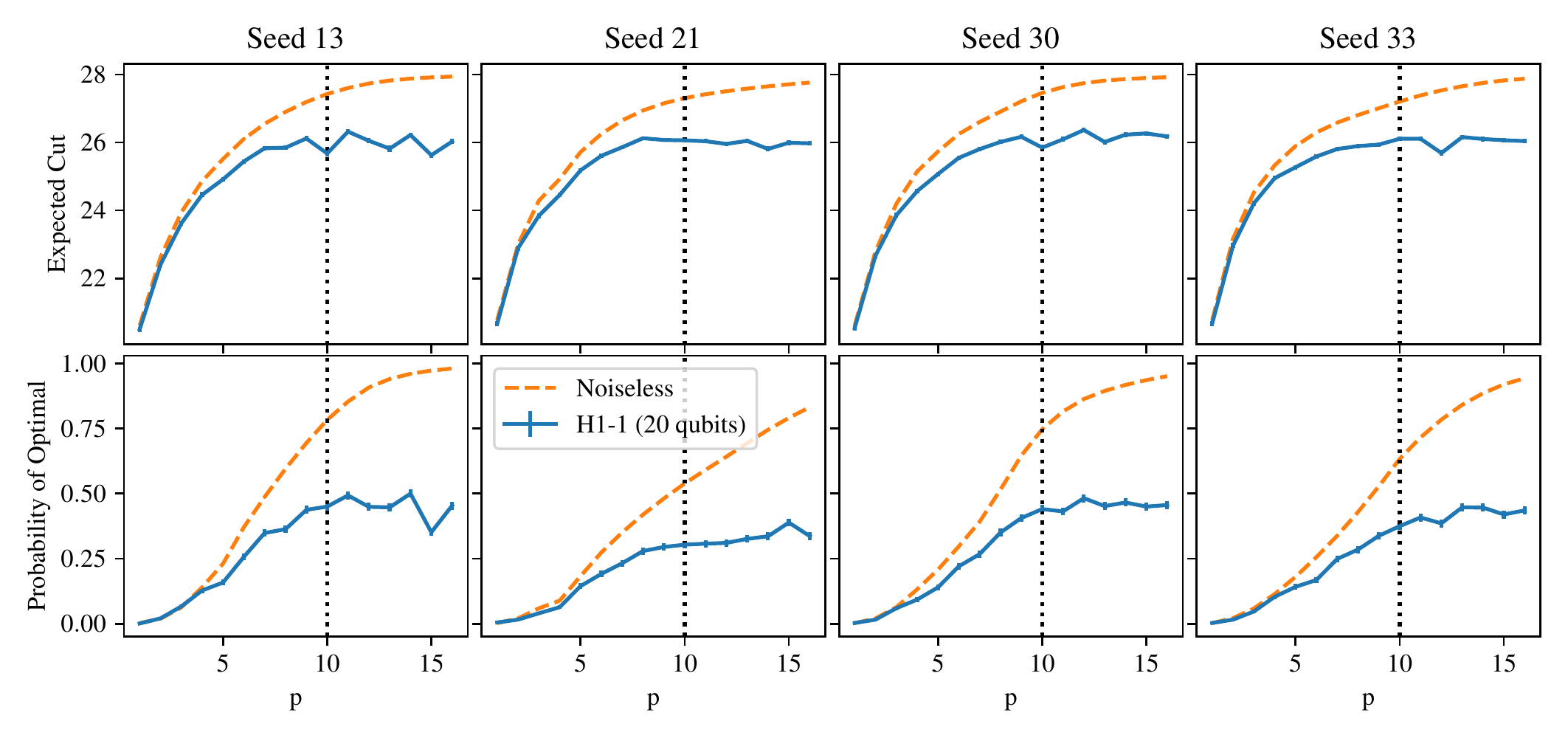}
    \caption{Values of the expected cut $\langle C\rangle$ (top), and probabilities $p^{\text{opt}}$ of obtaining the optimal solution (bottom) for all $N=20$ instances executed. QAOA parameters are optimized independently for each instance. The optimal cut value for all graphs is $28$ and the largest approximation ratio observed on hardware is $0.94$. The probability of obtaining the optimal solution increases monotonically for all instances until $p = 10$ (black vertical dotted line). Error bars show one standard error of the mean.}
    \label{fig:all_results}
\end{figure*}

\begin{figure*}
    \centering
    \includegraphics[width=\textwidth]{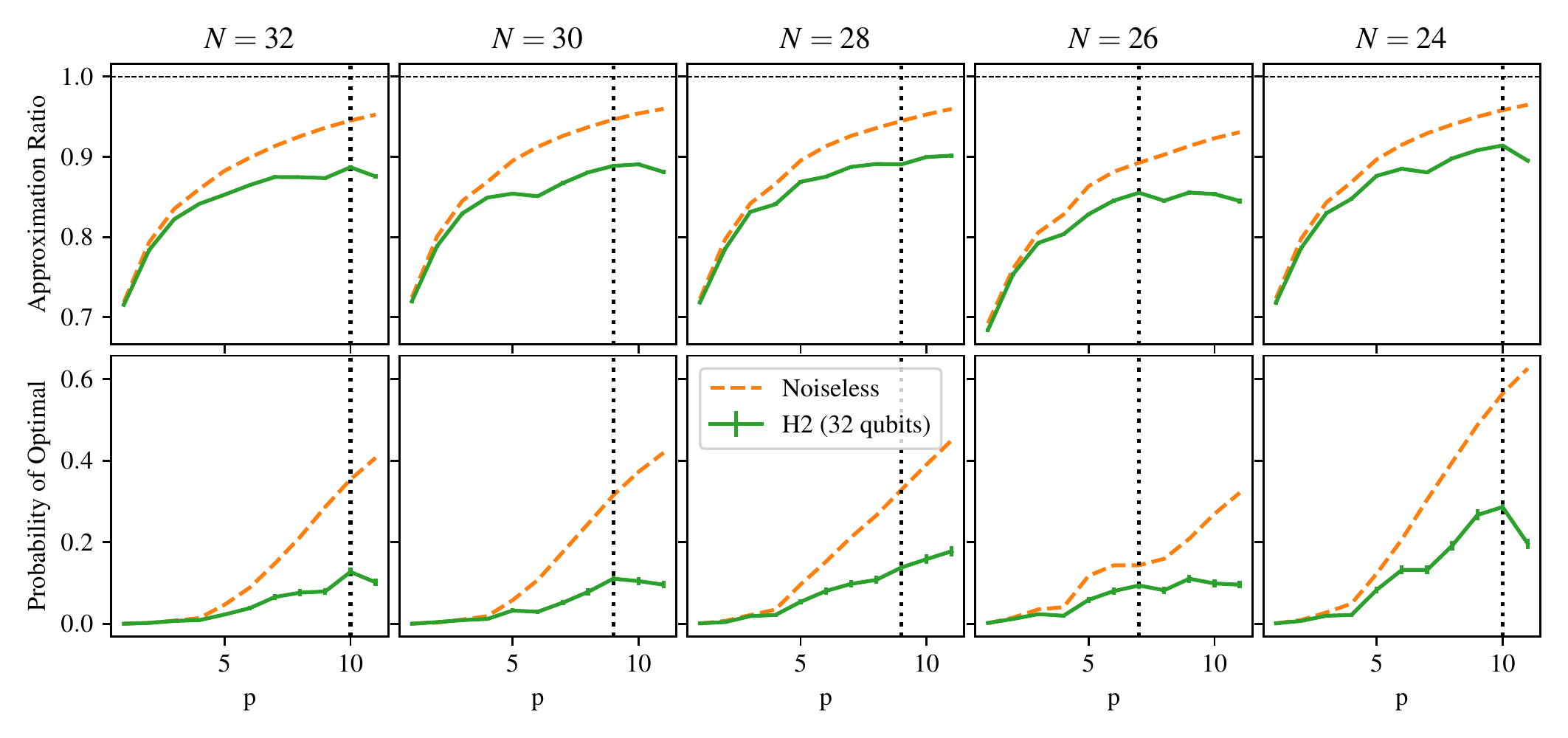}
    \caption{Approximation ratios (top), and probabilities $p^{\text{opt}}$ of obtaining the optimal solution (bottom) for all $N \geq 24$ instances executed. Fixed instance-independent QAOA parameters of Ref.~\cite{wurtz2021fixedangle} are used. Error bars show one standard error of the mean.}
    \label{fig:all_results_N_geq_24}
\end{figure*}

\paragraph{Experimental Setup} The instances are created by generating random 3-regular graphs and post-selecting on the value of optimal cut. For $N=28$, only the instances with optimal cut being at least $28$ are included. For higher $N$, instances with the largest value of optimal cut found within the sample are picked. We observe that the MaxCut problem for graphs with higher value of optimal cut is harder for QAOA, i.e., higher values of $p$ are required to approach optimal values of $\langle C\rangle$ and $p^{\text{opt}}$. Consequently, harder instances enable the demonstration of higher $N\cdot p$. In Figure~\ref{fig:all_results} and Table~\ref{tab:all_results}, each graph is identified by the random seed used to generate it. 

For $N=20$ instances, we optimize parameters $\bm\beta$, $\bm\gamma$ with respect to expected solution quality $\langle C (\bm \beta, \bm \gamma) \rangle$. For $p\leq 11$, we run one local optimization using COBYLA~\cite{Powell1994} initialized with fixed-angle parameters~\cite{wurtz2021fixedangle} obtained from QAOAKit~\cite{shaydulin2021qaoakit}. For $p>11$, we re-parameterize QAOA using the FOURIER scheme~\cite{zhou2020quantum} with $q=p$ and run one local optimization using COBYLA~\cite{Powell1994} initialized with parameters extrapolated from $p'=p-1$. For instances with $N\geq 24$, fixed parameters of Ref.~\cite{wurtz2021fixedangle} are used. The circuit with optimized parameters is compiled into the native gate set of the Quantinuum H1-1 device using the TKET transpiler~\cite{Sivarajah_2020}, with each $\zgate_i\zgate_j$ term in the cost operator $e^{-i \gamma_l C}$ implemented using one \texttt{ZZPhase} native two-qubit gate. Therefore, the two-qubit gate count for each circuit is exactly $p\cdot |E| = 3Np/2$. We remark that for the trapped-ion architecture of the H1-1 processor, the idling and crosstalk errors are low compared to the two-qubit gate errors. Combined with long coherence time, this leads to the success of the circuit execution being defined primarily by the two-qubit gate count, as opposed to the two-qubit gate depth. This is in contrast to superconducting architectures, for which the two-qubit gate depth is more predictive of performance than the two-qubit gate count due to the shorter coherence time and relatively higher crosstalk and idling errors.

\begin{table*}[t]
\scriptsize
\centering
\begin{subtable}[t]{0.48\textwidth}
\begin{tabular}{|c|c|c|c|c|}
\hline 
$p$ & $\langle C\rangle_{\text{H1-1}}$ & $\langle C\rangle_{\text{ex}}$  & $p^{\text{opt}}_{\text{H1-1}}$ & $p^{\text{opt}}_{\text{ex}}$ \\
\hline 
    $1$ & $20.500\pm 0.073$ & $20.610$ & $0.001\pm 0.001$ & $0.002$ \\
$2$ & $22.405\pm 0.073$ & $22.620$ & $0.021\pm 0.004$ & $0.019$ \\
$3$ & $23.620\pm 0.075$ & $23.944$ & $0.066\pm 0.008$ & $0.063$ \\
$4$ & $24.460\pm 0.074$ & $24.861$ & $0.128\pm 0.010$ & $0.140$ \\
$5$ & $24.913\pm 0.066$ & $25.512$ & $0.158\pm 0.011$ & $0.231$ \\
$6$ & $25.439\pm 0.067$ & $26.101$ & $0.257\pm 0.014$ & $0.371$ \\
$7$ & $25.829\pm 0.066$ & $26.547$ & $0.349\pm 0.015$ & $0.487$ \\
$8$ & $25.846\pm 0.068$ & $26.900$ & $0.363\pm 0.015$ & $0.595$ \\
$9$ & $26.116\pm 0.069$ & $27.189$ & $0.438\pm 0.016$ & $0.694$ \\
$10$ & $25.675\pm 0.097$ & $27.426$ & $0.450\pm 0.016$ & $0.784$ \\
$\mathbf{11}$ & $26.316\pm 0.065$ & $27.601$ & $\mathbf{0.493\pm 0.016}$ & $0.854$ \\
$12$ & $26.050\pm 0.072$ & $27.734$ & $0.449\pm 0.016$ & $0.908$ \\
$13$ & $25.816\pm 0.091$ & $27.821$ & $0.447\pm 0.016$ & $0.941$ \\
$14$ & $26.219\pm 0.068$ & $27.878$ & $0.500\pm 0.016$ & $0.961$ \\
$15$ & $25.625\pm 0.078$ & $27.914$ & $0.352\pm 0.015$ & $0.973$ \\
$16$ & $26.026\pm 0.071$ & $27.938$ & $0.454\pm 0.016$ & $0.981$ \\
\hline 
    \end{tabular}
    \caption{$N=20$, Seed $13$}
\end{subtable}
\begin{subtable}[t]{0.48\textwidth}
\begin{tabular}{|c|c|c|c|c|}
\hline 
$p$ & $\langle C\rangle_{\text{H1-1}}$ & $\langle C\rangle_{\text{ex}}$  & $p^{\text{opt}}_{\text{H1-1}}$ & $p^{\text{opt}}_{\text{ex}}$ \\
\hline 
$1$ & $20.663\pm 0.078$ & $20.774$ & $0.005\pm 0.002$ & $0.002$ \\
$2$ & $22.886\pm 0.073$ & $22.993$ & $0.016\pm 0.004$ & $0.020$ \\
$3$ & $23.843\pm 0.072$ & $24.286$ & $0.040\pm 0.006$ & $0.059$ \\
$4$ & $24.453\pm 0.068$ & $24.922$ & $0.063\pm 0.008$ & $0.089$ \\
$5$ & $25.177\pm 0.066$ & $25.715$ & $0.145\pm 0.011$ & $0.183$ \\
$6$ & $25.606\pm 0.063$ & $26.249$ & $0.191\pm 0.012$ & $0.274$ \\
$7$ & $25.858\pm 0.062$ & $26.649$ & $0.232\pm 0.013$ & $0.351$ \\
$8$ & $26.123\pm 0.059$ & $26.941$ & $0.279\pm 0.014$ & $0.419$ \\
$9$ & $26.071\pm 0.060$ & $27.150$ & $0.295\pm 0.014$ & $0.481$ \\
$10$ & $26.062\pm 0.064$ & $27.304$ & $0.304\pm 0.014$ & $0.539$ \\
$11$ & $26.035\pm 0.064$ & $27.418$ & $0.308\pm 0.014$ & $0.591$ \\
$12$ & $25.956\pm 0.065$ & $27.506$ & $0.311\pm 0.014$ & $0.642$ \\
$13$ & $26.045\pm 0.066$ & $27.584$ & $0.326\pm 0.015$ & $0.694$ \\
$14$ & $25.808\pm 0.073$ & $27.651$ & $0.336\pm 0.015$ & $0.745$ \\
$\mathbf{15}$ & $25.990\pm 0.075$ & $27.710$ & $\mathbf{0.389\pm 0.015}$ & $0.791$ \\
$16$ & $25.973\pm 0.065$ & $27.762$ & $0.337\pm 0.015$ & $0.832$ \\
\hline 
    \end{tabular}
    \caption{$N=20$, Seed $21$}
\end{subtable}\\
\vspace{0.1in}
\begin{subtable}[t]{0.48\textwidth}
\begin{tabular}{|c|c|c|c|c|}
\hline 
$p$ & $\langle C\rangle_{\text{H1-1}}$ & $\langle C\rangle_{\text{ex}}$  & $p^{\text{opt}}_{\text{H1-1}}$ & $p^{\text{opt}}_{\text{ex}}$ \\
\hline 
$1$ & $20.537\pm 0.076$ & $20.610$ & $0.003\pm 0.002$ & $0.002$ \\
$2$ & $22.671\pm 0.073$ & $22.797$ & $0.016\pm 0.004$ & $0.020$ \\
$3$ & $23.857\pm 0.072$ & $24.195$ & $0.060\pm 0.007$ & $0.064$ \\
$4$ & $24.568\pm 0.071$ & $25.146$ & $0.093\pm 0.009$ & $0.132$ \\
$5$ & $25.070\pm 0.067$ & $25.733$ & $0.140\pm 0.011$ & $0.208$ \\
$6$ & $25.547\pm 0.066$ & $26.248$ & $0.221\pm 0.013$ & $0.297$ \\
$7$ & $25.803\pm 0.064$ & $26.604$ & $0.268\pm 0.014$ & $0.392$ \\
$8$ & $26.015\pm 0.067$ & $26.906$ & $0.351\pm 0.015$ & $0.516$ \\
$9$ & $26.165\pm 0.065$ & $27.210$ & $0.406\pm 0.015$ & $0.646$ \\
$\mathbf{10}$ & $25.847\pm 0.091$ & $27.459$ & $\mathbf{0.440\pm 0.016}$ & $0.747$ \\
$11$ & $26.092\pm 0.075$ & $27.628$ & $0.432\pm 0.015$ & $0.816$ \\
$12$ & $26.365\pm 0.066$ & $27.745$ & $0.482\pm 0.016$ & $0.864$ \\
$13$ & $26.013\pm 0.081$ & $27.818$ & $0.452\pm 0.016$ & $0.895$ \\
$14$ & $26.229\pm 0.067$ & $27.865$ & $0.466\pm 0.016$ & $0.918$ \\
$15$ & $26.266\pm 0.064$ & $27.896$ & $0.450\pm 0.016$ & $0.936$ \\
$16$ & $26.175\pm 0.067$ & $27.919$ & $0.456\pm 0.016$ & $0.952$ \\
\hline 
    \end{tabular}
    \caption{$N=20$, Seed $30$}
\end{subtable}
\begin{subtable}[t]{0.48\textwidth}
\begin{tabular}{|c|c|c|c|c|}
\hline 
$p$ & $\langle C\rangle_{\text{H1-1}}$ & $\langle C\rangle_{\text{ex}}$  & $p^{\text{opt}}_{\text{H1-1}}$ & $p^{\text{opt}}_{\text{ex}}$ \\
\hline 
$1$ & $20.673\pm 0.080$ & $20.774$ & $0.003\pm 0.002$ & $0.002$ \\
$2$ & $22.963\pm 0.075$ & $23.167$ & $0.016\pm 0.004$ & $0.021$ \\
$3$ & $24.208\pm 0.067$ & $24.525$ & $0.047\pm 0.007$ & $0.059$ \\
$4$ & $24.948\pm 0.063$ & $25.327$ & $0.104\pm 0.010$ & $0.114$ \\
$5$ & $25.269\pm 0.064$ & $25.894$ & $0.142\pm 0.011$ & $0.180$ \\
$6$ & $25.585\pm 0.059$ & $26.293$ & $0.168\pm 0.012$ & $0.256$ \\
$7$ & $25.806\pm 0.060$ & $26.582$ & $0.249\pm 0.014$ & $0.338$ \\
$8$ & $25.892\pm 0.061$ & $26.806$ & $0.285\pm 0.014$ & $0.429$ \\
$9$ & $25.936\pm 0.068$ & $27.006$ & $0.338\pm 0.015$ & $0.528$ \\
$10$ & $26.107\pm 0.063$ & $27.204$ & $0.375\pm 0.015$ & $0.633$ \\
$\mathbf{11}$ & $26.108\pm 0.068$ & $27.384$ & $\mathbf{0.408\pm 0.015}$ & $0.716$ \\
$12$ & $25.688\pm 0.086$ & $27.532$ & $0.386\pm 0.015$ & $0.784$ \\
$13$ & $26.158\pm 0.066$ & $27.653$ & $0.447\pm 0.016$ & $0.841$ \\
$14$ & $26.100\pm 0.071$ & $27.751$ & $0.446\pm 0.016$ & $0.886$ \\
$15$ & $26.064\pm 0.067$ & $27.824$ & $0.420\pm 0.015$ & $0.920$ \\
$16$ & $26.039\pm 0.068$ & $27.875$ & $0.436\pm 0.015$ & $0.944$ \\
\hline 
    \end{tabular}
    \caption{$N=20$, Seed $33$}
\end{subtable}
    \caption{Expected values of cut ($\langle C\rangle$) and probabilities of obtaining the optimal solution ($p^{\text{opt}}$) with the corresponding standard error of the mean for all $N=20$ instances executed on H1-1. $\langle C\rangle_{\text{ex}}$ and $p^{\text{opt}}_{\text{ex}}$ are obtained in exact noiseless simulation. Note that for most instances $p^{\text{opt}}$ continues to improve beyond $p=10$. The last $p$ for which the $p^{\text{opt}}$ is still increasing is highlighted in bold.}
    \label{tab:all_results}
\end{table*}

\begin{table*}[t]
\scriptsize
\centering
\begin{subtable}[t]{0.48\textwidth}
\begin{tabular}{|c|c|c|c|c|}
\hline 
$p$ & $\langle C\rangle_{\text{H2}}$ & $\langle C\rangle_{\text{ex}}$  & $p^{\text{opt}}_{\text{H2}}$ & $p^{\text{opt}}_{\text{ex}}$ \\
\hline 
$1$ & $32.910\pm 0.097$ & $33.071$ & $0.000\pm 0.000$ & $0.000$ \\
$2$ & $36.043\pm 0.090$ & $36.489$ & $0.002\pm 0.001$ & $0.002$ \\
$3$ & $37.823\pm 0.088$ & $38.413$ & $0.007\pm 0.003$ & $0.007$ \\
$4$ & $38.705\pm 0.084$ & $39.556$ & $0.009\pm 0.003$ & $0.014$ \\
$5$ & $39.228\pm 0.089$ & $40.599$ & $0.022\pm 0.005$ & $0.046$ \\
$6$ & $39.780\pm 0.091$ & $41.350$ & $0.038\pm 0.006$ & $0.088$ \\
$7$ & $40.242\pm 0.102$ & $42.018$ & $0.065\pm 0.008$ & $0.148$ \\
$8$ & $40.230\pm 0.101$ & $42.576$ & $0.076\pm 0.008$ & $0.213$ \\
$9$ & $40.181\pm 0.111$ & $43.067$ & $0.079\pm 0.008$ & $0.286$ \\
$\mathbf{10}$ & $40.800\pm 0.108$ & $43.478$ & $\mathbf{0.127\pm 0.010}$ & $0.354$ \\
$11$ & $40.272\pm 0.117$ & $43.819$ & $0.102\pm 0.009$ & $0.407$ \\
\hline 
    \end{tabular}
    \caption{$N=32$}
\end{subtable}
\begin{subtable}[t]{0.48\textwidth}
\begin{tabular}{|c|c|c|c|c|}
\hline 
$p$ & $\langle C\rangle_{\text{H2}}$ & $\langle C\rangle_{\text{ex}}$  & $p^{\text{opt}}_{\text{H2}}$ & $p^{\text{opt}}_{\text{ex}}$ \\
\hline 
$1$ & $30.961\pm 0.097$ & $31.160$ & $0.000\pm 0.000$ & $0.000$ \\
$2$ & $33.915\pm 0.088$ & $34.424$ & $0.004\pm 0.002$ & $0.003$ \\
$3$ & $35.662\pm 0.094$ & $36.335$ & $0.009\pm 0.003$ & $0.010$ \\
$4$ & $36.515\pm 0.082$ & $37.357$ & $0.012\pm 0.003$ & $0.019$ \\
$5$ & $36.725\pm 0.101$ & $38.486$ & $0.032\pm 0.006$ & $0.058$ \\
$6$ & $36.587\pm 0.101$ & $39.237$ & $0.029\pm 0.005$ & $0.107$ \\
$7$ & $37.292\pm 0.102$ & $39.823$ & $0.052\pm 0.007$ & $0.176$ \\
$8$ & $37.865\pm 0.097$ & $40.289$ & $0.078\pm 0.008$ & $0.245$ \\
$\mathbf{9}$ & $38.213\pm 0.101$ & $40.694$ & $\mathbf{0.110\pm 0.010}$ & $0.315$ \\
$10$ & $38.297\pm 0.101$ & $41.021$ & $0.104\pm 0.010$ & $0.372$ \\
$11$ & $37.887\pm 0.106$ & $41.275$ & $0.096\pm 0.009$ & $0.419$ \\
\hline 
    \end{tabular}
    \caption{$N=30$}
\end{subtable}\\
\vspace{0.1in}
\begin{subtable}[t]{0.48\textwidth}
\begin{tabular}{|c|c|c|c|c|}
\hline 
$p$ & $\langle C\rangle_{\text{H2}}$ & $\langle C\rangle_{\text{ex}}$  & $p^{\text{opt}}_{\text{H2}}$ & $p^{\text{opt}}_{\text{ex}}$ \\
\hline 
$1$ & $28.734\pm 0.090$ & $28.916$ & $0.001\pm 0.001$ & $0.000$ \\
$2$ & $31.405\pm 0.088$ & $31.877$ & $0.004\pm 0.002$ & $0.007$ \\
$3$ & $33.253\pm 0.086$ & $33.670$ & $0.019\pm 0.004$ & $0.021$ \\
$4$ & $33.635\pm 0.083$ & $34.644$ & $0.021\pm 0.005$ & $0.035$ \\
$5$ & $34.754\pm 0.082$ & $35.811$ & $0.054\pm 0.007$ & $0.096$ \\
$6$ & $35.006\pm 0.086$ & $36.528$ & $0.080\pm 0.008$ & $0.152$ \\
$7$ & $35.495\pm 0.086$ & $37.042$ & $0.098\pm 0.009$ & $0.212$ \\
$8$ & $35.642\pm 0.086$ & $37.430$ & $0.107\pm 0.010$ & $0.265$ \\
$\mathbf{9}$ & $35.622\pm 0.099$ & $37.786$ & $\mathbf{0.138\pm 0.011}$ & $0.328$ \\
$10$ & $35.989\pm 0.093$ & $38.110$ & $0.158\pm 0.011$ & $0.390$ \\
$11$ & $36.060\pm 0.100$ & $38.388$ & $0.178\pm 0.012$ & $0.450$ \\
\hline 
    \end{tabular}
    \caption{$N=28$}
\end{subtable}
\begin{subtable}[t]{0.48\textwidth}
\begin{tabular}{|c|c|c|c|c|}
\hline 
$p$ & $\langle C\rangle_{\text{H2}}$ & $\langle C\rangle_{\text{ex}}$  & $p^{\text{opt}}_{\text{H2}}$ & $p^{\text{opt}}_{\text{ex}}$ \\
\hline 
$1$ & $26.683\pm 0.088$ & $27.006$ & $0.002\pm 0.001$ & $0.001$ \\
$2$ & $29.394\pm 0.094$ & $29.668$ & $0.012\pm 0.003$ & $0.015$ \\
$3$ & $30.917\pm 0.092$ & $31.420$ & $0.023\pm 0.005$ & $0.035$ \\
$4$ & $31.335\pm 0.088$ & $32.292$ & $0.020\pm 0.004$ & $0.040$ \\
$5$ & $32.312\pm 0.092$ & $33.672$ & $0.059\pm 0.007$ & $0.118$ \\
$6$ & $32.971\pm 0.100$ & $34.375$ & $0.080\pm 0.008$ & $0.143$ \\
$\mathbf{7}$ & $33.352\pm 0.094$ & $34.811$ & $\mathbf{0.094\pm 0.009}$ & $0.143$ \\
$8$ & $32.966\pm 0.104$ & $35.204$ & $0.082\pm 0.009$ & $0.159$ \\
$9$ & $33.363\pm 0.111$ & $35.626$ & $0.110\pm 0.010$ & $0.209$ \\
$10$ & $33.288\pm 0.110$ & $36.004$ & $0.099\pm 0.009$ & $0.270$ \\
$11$ & $32.954\pm 0.115$ & $36.295$ & $0.096\pm 0.009$ & $0.321$ \\
\hline 
    \end{tabular}
    \caption{$N=26$}
\end{subtable}\\
\vspace{0.1in}
\begin{subtable}[t]{0.48\textwidth}
\begin{tabular}{|c|c|c|c|c|}
\hline 
$p$ & $\langle C\rangle_{\text{H2}}$ & $\langle C\rangle_{\text{ex}}$  & $p^{\text{opt}}_{\text{H2}}$ & $p^{\text{opt}}_{\text{ex}}$ \\
\hline 
$1$ & $24.418\pm 0.085$ & $24.595$ & $0.001\pm 0.001$ & $0.001$ \\
$2$ & $26.734\pm 0.082$ & $27.122$ & $0.007\pm 0.003$ & $0.009$ \\
$3$ & $28.208\pm 0.078$ & $28.663$ & $0.020\pm 0.004$ & $0.028$ \\
$4$ & $28.818\pm 0.070$ & $29.529$ & $0.021\pm 0.005$ & $0.049$ \\
$5$ & $29.794\pm 0.071$ & $30.489$ & $0.083\pm 0.009$ & $0.123$ \\
$6$ & $30.092\pm 0.075$ & $31.111$ & $0.132\pm 0.011$ & $0.206$ \\
$7$ & $29.940\pm 0.084$ & $31.595$ & $0.132\pm 0.011$ & $0.304$ \\
$8$ & $30.530\pm 0.078$ & $31.968$ & $0.191\pm 0.012$ & $0.396$ \\
$9$ & $30.880\pm 0.082$ & $32.296$ & $0.267\pm 0.014$ & $0.487$ \\
$\mathbf{10}$ & $31.079\pm 0.079$ & $32.577$ & $\mathbf{0.286\pm 0.014}$ & $0.564$ \\
$11$ & $30.434\pm 0.083$ & $32.806$ & $0.196\pm 0.012$ & $0.626$ \\
\hline 
    \end{tabular}
    \caption{$N = 24$}
\end{subtable}
    \caption{Expected values of cut ($\langle C\rangle$) and probabilities of obtaining the optimal solution ($p^{\text{opt}}$) with the corresponding standard error of the mean for all $N\geq 24$ instances executed on H2. $\langle C\rangle_{\text{ex}}$ and $p^{\text{opt}}_{\text{ex}}$ are obtained in exact noiseless simulation. Note that for most instances $p^{\text{opt}}$ continues to improve beyond $p=10$. The last $p$ for which the $p^{\text{opt}}$ is still increasing is highlighted in bold.}
    \label{tab:all_results_N_geq_24}
\end{table*}

\paragraph{QAOA on the H1-1 and H2 Trapped-Ion Processors} We present the results obtained on the Quantinuum H1-1 quantum processor in Figure~\ref{fig:all_results} and on the H2 in Figure~\ref{fig:all_results_N_geq_24}. Table~\ref{tab:all_results} lists the complete results for Figure~\ref{fig:all_results} and Table~\ref{tab:all_results_N_geq_24} for Figure~\ref{fig:all_results_N_geq_24}. The executed circuits and the raw data obtained from the device are available at \url{https://doi.org/10.5281/zenodo.7689982}. We use the probability $p^{\text{opt}}$ of obtaining the optimal cut as the measure of solution quality for the purposes of defining a ``successful execution''. 

For $N=20$ experiments with instance-specific optimized QAOA parameters, we observe the probability $p^{\text{opt}}$ of obtaining the optimal cut grows monotonically up to $p=10$ that for all four graphs considered. This indicates that the experiments consistently succeed at $N\cdot p = 200$, which corresponds to $300$ native two-qubit \texttt{ZZPhase} gates. We obtain approximation ratios of up to $0.94$. We note that $p^{\text{opt}}$ continues to grow for higher $p$ compared to the expected solution quality $\langle C\rangle$, which for some instances stops increasing at $p=9$. We conjecture that this difference is due to higher $p$ required to saturate the success probability in the noiseless case. This conjecture is supported by the observation that the highest $N\cdot p$ is achieved with instances for which $p^{\text{opt}}_{\text{ex}}$ grows the slowest. For example, the instance labelled ``Seed 21'' achieves $N\cdot p = 300$ and $p^{\text{opt}}_{\text{ex}}=0.791$ at $p=15$. In contrast, the same $p^{\text{opt}}_{\text{ex}}$ is reached at $p=11$ for ``Seed 13'' and ``Seed 30'', and $N\cdot p$ is correspondingly lower.

For $N\geq 24$ experiments with fixed QAOA parameters, we observe a less consistent performance. For example, for $N=32$ the value of $p^{\text{opt}}$ grows for $p$ up to 10, whereas for $N=26$ the monotonic improvement continues only up to $p=7$. This is due to the fixed QAOA parameters providing non-smooth improvement in $p^{\text{opt}}$ for $N=26$ instance, as evidences by the exact simulation presented in Fig.~\ref{fig:all_results_N_geq_24}. To the best of our knowledge, this note reports the largest QAOA execution in terms of $N\cdot p$ to date. We refer the interested reader to Table~2 in Ref.~\cite{Niroula2022} for an overview of previous state-of-the-art demonstrations. We remark that alternative definitions of QAOA layer may lead to smaller circuits that have higher $N\cdot p$. For example, in the recently introduced mixer-phaser ansatz~\cite{LaRose2022}, each ``layer'' can be implemented using only $3$ $\cnotgate$s. If ``layer''  is to be redefined in this way, our results indicate that $N\cdot p \geq 2000$ would become within reach for current hardware. However, in this note we focus on the original QAOA definition.

 We further note that neither $\langle C\rangle$ nor $p^{\text{opt}}$ measurably decrease as more layers are added. This is in contrast to previous results on superconducting~\cite{harrigan2021quantum,Lacroix_2020} and trapped-ion~\cite{2306.03238} quantum processors, where at some point additional layers begin to introduce an amount of noise sufficient to significantly decrease the expected solution quality. Combined, these observations suggest that the current experiment is not yet limited by the gate fidelity of the device. Higher values of $N\cdot p$ may be achieved at current gate fidelities either by considering harder problems (higher $p$) or by loading more ions into the trap (higher $N$).

\section*{Acknowledgements}

The authors thank Dylan Herman, Changhao Li, Yue Sun and other members of the Global Technology Applied Research center of JPMorgan Chase for helpful discussions and providing feedback on the manuscript.

\bibliographystyle{IEEEtran}
\bibliography{sample}

\section*{Disclaimer}
This paper was prepared for informational purposes by the Global Technology Applied Research center of JPMorgan Chase \& Co. This paper is not a product of the Research Department of JPMorgan Chase \& Co. or its affiliates. Neither JPMorgan Chase \& Co. nor any of its affiliates makes any explicit or implied representation or warranty and none of them accept any liability in connection with this paper, including, without limitation, with respect to the completeness, accuracy, or reliability of the information contained herein and the potential legal, compliance, tax, or accounting effects thereof. This document is not intended as investment research or investment advice, or as a recommendation, offer, or solicitation for the purchase or sale of any security, financial instrument, financial product or service, or to be used in any way for evaluating the merits of participating in any transaction.
\end{document}